\documentclass[runningheads]{svmult}

\usepackage{makeidx}   
\usepackage{graphicx}  
\usepackage{subeqnar}  
\usepackage{multicol}  
\usepackage{physprbb}  
\makeindex             

\begin{document}
\title*{Abundance trends in the thin and thick disks}
\toctitle{Abundance trends in the thin and thick disks}
%
%
\titlerunning{Abundance trends in the thin and thick disks}
%
\author{Sofia Feltzing
}
\authorrunning{Sofia Feltzing}
%
%
\institute{Lund Observatory, Box 43, SE-221 00 Lund, Sweden}

\maketitle              

\begin{abstract}
The Milky Way harbours two disks that appear distinct concerning scale-heights,
kinematics, and elemental abundance patterns. 
Recent years have seen a surge of 
studies of the elemental abundance trends in the disks using high
resolution spectroscopy. Here I will review and discuss the currently available
data.  Special focus
will also be put on how we define stars to be members
of either disk, and how current models of galaxy formation favour that 
thick disks are formed from several accreted bodies. 
The ability for the stellar
abundance trends to test such predictions are discussed.
\end{abstract}

\section{Current observational knowledge about the thick disk}

During the last decade there has been an increasing interest in trying
to establish the chemical abundance trends for the thick disk in the Milky Way.
 Some of 
these have included stars from both the thin and the thick disk. 
These are the most useful ones as 
they perform differential studies between the two disks. This means
that shortcomings in the abundance analyses are, to first order, canceled.
This is especially true if the stars are selected to span a small 
range in effective temperature
and surface gravity. Then any resulting difference, at a given [Fe/H], between
the two disks may be regarded as real.  

Recent important observational studies include the following (sorted 
according to the stellar samples): 
{\sl Differential studies}: Fuhrmann, 1998, \cite{feltzing:fuhrmann98} and
 2004, \cite{feltzing:fuhrmann04}, Chen et al., 2000, \cite{feltzing:chen}, 
Mashonkina et al., 2003, \cite{feltzing:mashonkina}, Gratton et al., 2003, 
\cite{feltzing:gratton03}, Bensby et al., 2003, \cite{feltzing:bensby03}  and  2004, 
\cite{feltzing:bensby04a}, Mishenina et al., 2004,\cite{feltzing:mishenina}
{\sl Thick disk only:} Prochaska et al., 2000, \cite{feltzing:prochaska}
{\sl Thin disk only:} Reddy et al., 2003, \cite{feltzing:reddy}, Allende 
Prieto et al., 2004,
 \cite{feltzing:allende}

Although the various studies cited above take different approaches to defining
their samples and though some of them are only concerned with one of the disks,
there is agreement on the following: 
\begin{itemize}

\item The thick disk is, at a given [Fe/H], 
more enhanced in the $\alpha$-elements than the thin disk 

\item The abundance trends in the thin disk is a gentle slope 

\item  The solar neighbourhood thick disk stars that 
have been studied are all old

\end{itemize}

Other issues are less clear cut.
There is some agreement on the following issues (e.g. {\cite{feltzing:mashonkina}, 
\cite{feltzing:bensby03}, and Fig. \ref{feltzing:euo} \& \ref{feltzing:mg}).
\begin{itemize}
\item The thick disk shows evidence of extensive star formation

\item The thick disk shows evidence of pollution from SNe\,Ia

\item The thick disk shows evidence of pollution from AGB stars
\end{itemize}

 In addition to the indication of pollution from SNe\,Ia (i.e. the ``knee'') 
found in the [Mg/Fe]
and [O/Fe]  trends,for kinematically selected thick disk samples,
 there is also
evidence that  thick disk stars well above the galactic plane show the 
same trend. In Fig. \ref{feltzing:mg} we show the first results
of a study of ``in situ'' metal-rich thick disk stars. It appears that there
might be a downward trend in [Mg/Fe] as a function of [Fe/H] also 
well above the thin disk. This must, however, be further established
before any firm conclusions can be drawn.

There is no agreement on the following two issues (\cite{feltzing:gratton03}, \cite{feltzing:fuhrmann04}, 
\cite{feltzing:bensby03}, and \cite{feltzing:bensby04b}): 
\begin{itemize}
\item The thick disk extends to [Fe/H] = 0 
\item A true age-gap between the end of star formation in the thick disk and the onset 
of star formation in the thin disk
\end{itemize}

\begin{itemize}
\item Finally: Age-metallicity relation in the thick disk -- this is a 
very tentative statement and, to our knowledge, there is only one study that claims the 
{\sl possibility}
of such an age-metallicity relation (\cite{feltzing:bensby04b}). This would suggest an extended 
period of star formation in the thick disk
\end{itemize}

\section{Ways to define a thick disk star}

\textbf{Kinematics} Selection is clean, even though not trivial, as the velocities
of the thick disk overlaps those of the thin. 
It is easy to compare selection criteria between different studies.

\noindent
\textbf{Position} Sufficiently high above the galactic plane the 
thick disk dominates over both the thin disk and the halo. 
Thus if we can pick stars ~1000-1500 pc above the plane we 
have a good chance of picking a thick disk star.
But, these stars are faint due to their large distance (typically
V$\sim 16$)  and hence difficult to study.

\noindent
\textbf{[Fe/H] and [$\alpha$/Fe]} The metallicity distribution 
functions of the various stellar populations present in the solar
 neighbourhood  overlap greatly and hence 
metallicity on its own is not useful for classifying stars. 
The enhancements of $\alpha$-elements (e.g. O 
and Si) are sometimes used to identify the various populations. 
There is currently general agreement that the thick disk
is enhanced in these elements relative to the thin disk.
This should be regarded as something we are looking for rather than
prior knowledge that can be used to select stars for further study.

\noindent
\textbf{Age} Stellar ages can be derived using isochrones. 
However, such data can not be used to select stars for studies 
of the thick disk as this must imply a prior knowledge 
of the age of the stellar population in this disk. This is
 knowledge that we seek, but do not have.

Given that we in principle can never select a sample of local thick disk
stars that is guaranteed to not contain any intervening thin disk stars,
I would argue that we should keep the selection schemes as simple and as
transparent as possible. In this sense the simplest and most robust 
selection is based on the kinematics of the stars. This is also the least model
dependent method. Of course, should positions be 
available, i.e. height over the galactic plane, these could, and should, 
be used.

\begin{figure}
\begin{center}
\includegraphics[width=7cm,bb=18 144 592 475,clip]{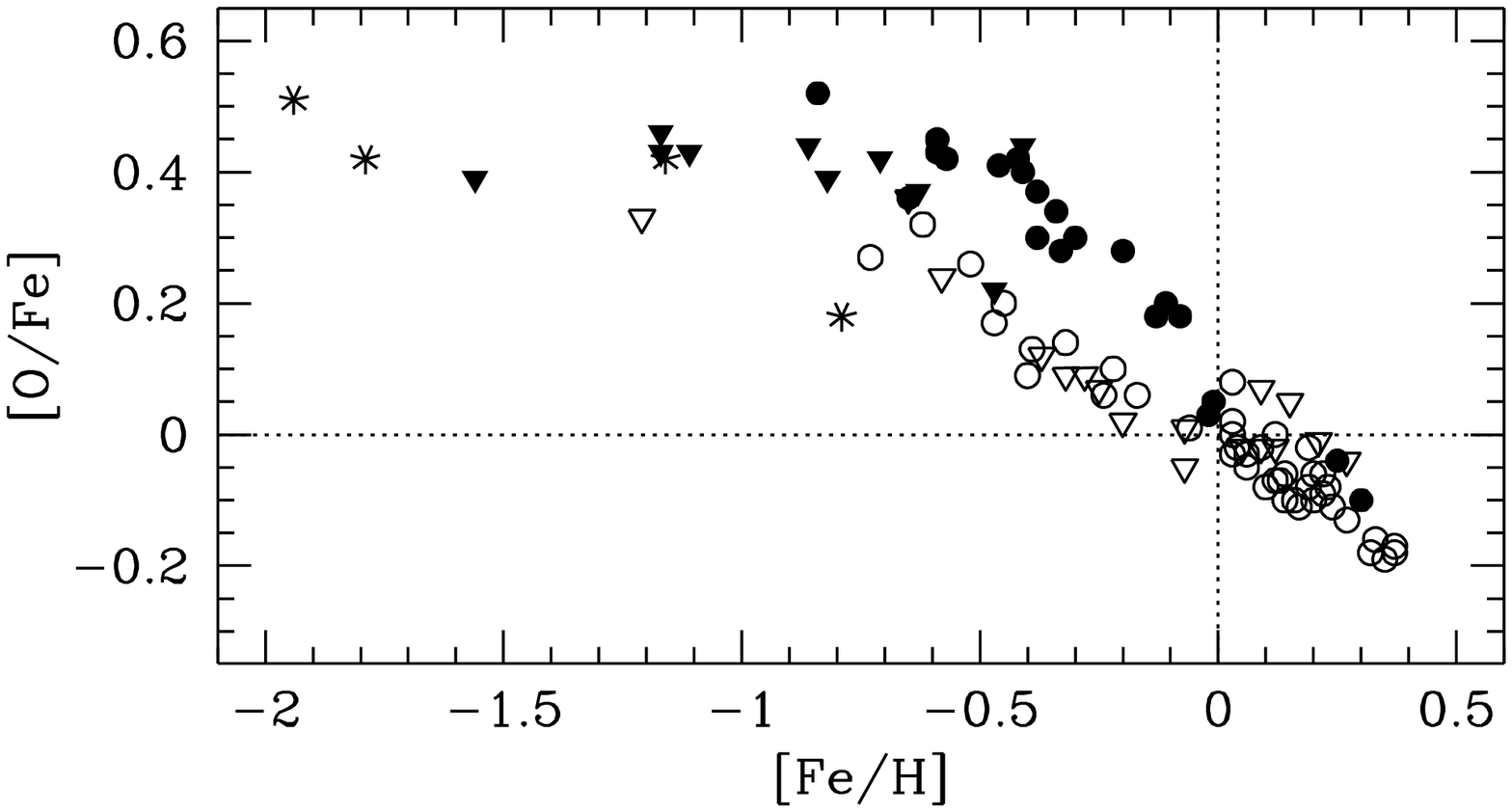}
\includegraphics[width=4.5cm]{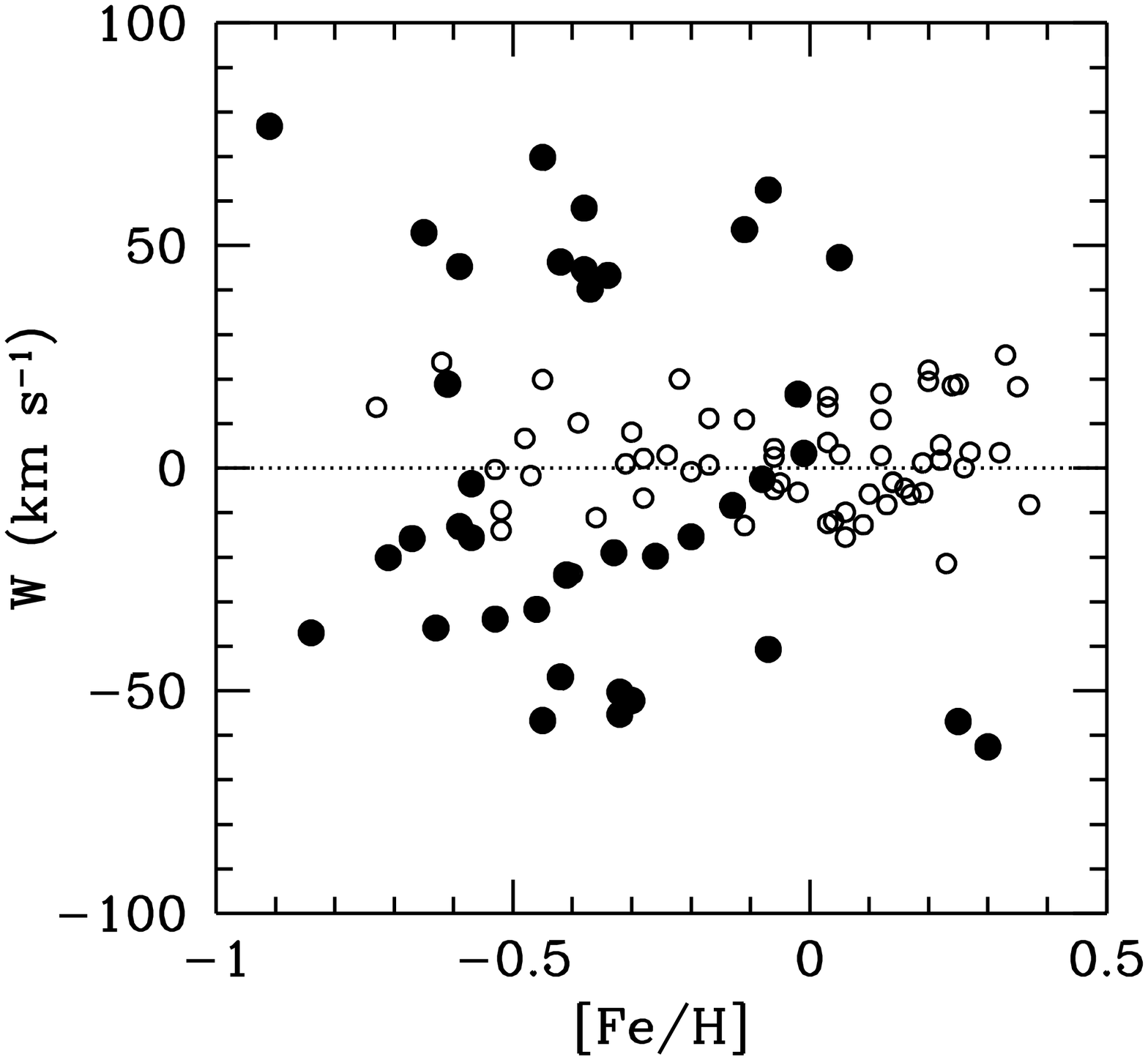}
\end{center}
\caption[]{{\bf Left panel:} Demonstration that SNe\,Ia have contributed to
the chemical enrichment 
for stars that are kinematically
selected to trace the thick disk (i.e. ``knee'' in the trend for [O/Fe]). Figure taken from 
\cite{feltzing:bensby04a}. Thick disk stars are represented by filled symbols (circles from
\cite{feltzing:bensby04a} and triangles from \cite{feltzing:nissen}), thin disk by open symbols (circles from
\cite{feltzing:bensby04a} and triangles from \cite{feltzing:nissen}), and 
halo stars by $\ast$ (from \cite{feltzing:nissen}). {\bf Right panel:} Shows the $W$-velocity as 
a function of [Fe/H] for stars kinematically selected to be thin (open circles) or thick
(filled circles) disk stars. Data taken from \cite{feltzing:bensby03} and Bensby et al. submitted.
This plot establishes that  metal-rich stars can also 
have high velocities perpendicular to the 
plane (sometimes taken as an evidence for the star to belong to the thick disk) }
\label{feltzing:euo}
\end{figure}

\begin{figure}
\begin{center}
\includegraphics[width=7cm,angle=-90]{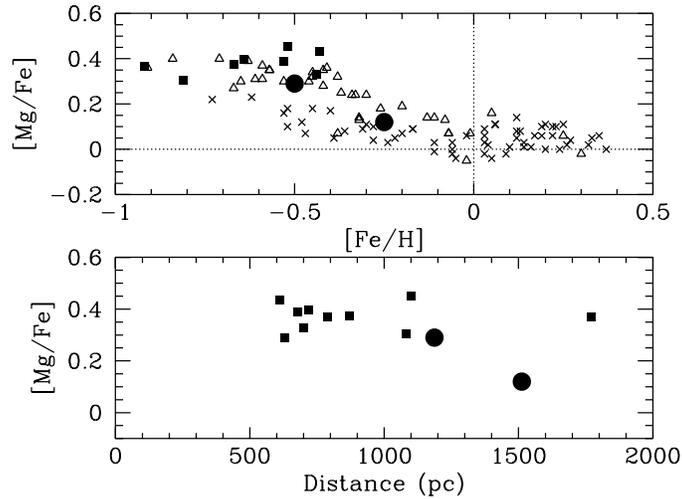}
\end{center}
\caption[]{{\bf Upper panel:}
First results from a study of five ``in situ'' thick disk stars (Feltzing et al. in prep.). 
The Mg abundance for two of them are indicated by large $\bullet$
 while data from \cite{feltzing:bensby03} and Bensby et al. in prep. are shown
with $\triangle$ for thick disks stars and $\times$ for thin disk stars.
The thick disk stars from \cite{feltzing:prochaska} are shown as filled squares. 
{\bf Lower panel:} [Mg/Fe] vs the distance above the galactic plane for 
the two stars from Feltzing et al. in prep. and the thick disk stars from \cite{feltzing:prochaska}}
\label{feltzing:mg}
\end{figure}

\begin{figure}
\begin{center}
\includegraphics[width=8cm]{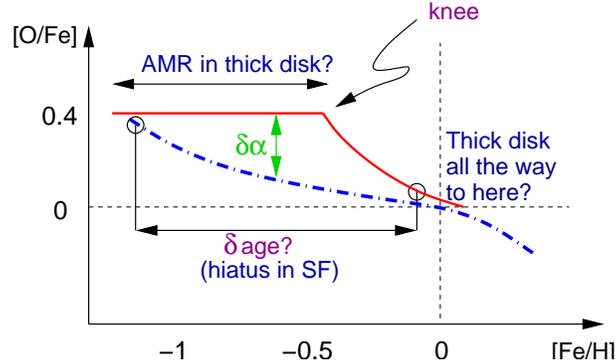}
\end{center}
\caption[]{Summary of current observational knowledge about the thin and the thick 
disks in the Milky Way. The two trends for [O/Fe] vs [Fe/H] 
are depicted in blue, dashed line (thin disk) and red (thick disk).
The most debated issues are marked in blue (i.e. thick disk all the way to
[Fe/H]=0, hiatus in star formation, SF, and AMR in thick disk) and
topics of some debate in purple 
(``knee'' and $\delta$age between various sub-populations,
such as
 $\delta$age between the youngest thin disk and the oldest thick disk). The one issue all 
agree upon, the $\alpha$-enhancement, is indicated in green (light grey)
}
\label{feltzing:summary}
\end{figure}

\section{Thick disks in other galaxies and how to form them}

\subsection{Thick disks in other galaxies and disks and old stars at high z}

Thick disks are not unique to the Milky Way.
Thick disks are seen in many spiral and lenticular galaxies, see e.g. 
\cite{feltzing:pohlen}, and in galaxies in merging 
environments, see e.g. \cite{feltzing:schwarzkopf}.
Some, \cite{feltzing:dalcanton}, even suggest that all spiral galaxies have 
thick disks.  It is an important observational
task to verify and extend these findings. 

It is also interesting to note that 
solar metallicities are reported for z=2, e.g. \cite{feltzing:shapley},
and that disks of old stars have been found at redshifts as high
as  z=2.5, see e.g. \cite{feltzing:stockton}. These types of findings indicate 
that indeed the formation of the thick disk in our galaxy might 
have happened well in the past.

\subsection{Ways to form thick disks}
\label{feltzing:models}

Meanwhile, the quest to find formation scenarios for the  thick
disks is ongoing. Earlier studies focused on fast and slow monolithic
collapse and various scenarios for increasing the velocity dispersion in
a pre-existing thin disk. The latter was envisaged to be able to happen in
two distinct ways; either through the general diffusion of orbits and stars
being scattered by molecular clouds and other stars or by a 
violent encounter and merging with a smaller galaxy. A good summary and 
discussion of why some of these scenarios do not work (e.g. diffusion of orbits
and slow collapse) while others (such as fast collapse) is still viable 
can be found in \cite{feltzing:gilmore}. Current models of 
galaxy formation envisages that a galaxy formed from many building blocks.
These types of models have only recently become detailed enough that 
we can attempt to compare them in greater detail with the data for the
Milky Way, e.g. \cite{feltzing:abadi} and \cite{feltzing:brooks}.

I'll review here three different scenarios that still appear viable
(based on the currently available data for the Milky Way,
see summary in Fig. \ref{feltzing:summary}). For each example
I have chosen one or two references that have done detailed models as 
illustrations to compare the observational
data to. This is not an exhaustive account for all the possibilities
within each scenario but it gives a flavour of the types of comparisons
we ought to make and, hopefully, it also illustrates the shortcomings
both of the observed as well as simulated data.

\medskip
\noindent
{\bf In an early phase with enhanced star formation (\cite{feltzing:burkert})}

\noindent
{\it Predictions:} no abundance gradients the thick disk; thin disk stars will 
all have [Fe/H] larger than found in the thick disk; period of star 
formation in thick disk was short $<$ 1 billion years; thin disk always younger 
than thick disk

\noindent
{\it Observations:}} no vertical gradients;
 metallicity distributions for the disks overlap;
 star formation in thick disk includes SNe\,Ia and AGB;
 star formation in thick disk probably $> 1$ billion years

\noindent
{\it Conclusions:} probably not a viable scenario 
(if an age-metallicity relation
 in thick disk is established this formation scenario is in serious trouble)

\medskip
\noindent
{\bf As the result if violent heating of a pre-existing thin disk (\cite{feltzing:quinn})}

 \noindent
{\it Predictions:} infalling satellite probably has to be large;
 thin disk younger than thick disk and an age gap between the disks;
 likely that the abundance trends differ;
 gradients will be preserved (if they exist in pre-existing thin disk);
 star formation in original thin disk could be as long as needed (e.g. to
create the ``knee'' in the $\alpha$-element trends)

 \noindent
{\it Observations:} no vertical gradients;
 abundance trends differ; star formation in thick disk includes SNe\,Ia and AGB

 \noindent
{\it Conclusions:} quite possible, no obvious problems with currently available observations

\medskip
\noindent
{\bf Are made from (largish) satellites that accrete over time (\cite{feltzing:abadi}, \cite{feltzing:brooks})}

 \noindent
{\it Predictions:} we will only see the abundance trend in 
the one satellite that finally settles in a torus at our R;
 the abundance trends will differ depending on the
 potential well of the satellite, it must be largish and not a dSph; 
no age gap between thin and thick disk as thin disk is accreted also 
at early times

 \noindent
{\it Observations:} tight trends in kinematically selected samples;
 so far all studied thick disk stars are older than thin disk stars

 \noindent
{\it Conclusions:} satellite must be fairly large;
 only one satellite contributes at the solar radius;
 the age structure remains a problem (i.e. that very old thin disk stars
are predicted and not observed, so far?)

\section{Summary}

There are a number of important observational facts that we now have
established for the stellar population of solar neighbourhood stars
that have kinematics that are typical of the galactic thick disk.
Most importantly it is now convincingly shown by independent studies 
that the abundance trends for kinematically selected samples differ;
e.g. the thick disk is more enhanced in $\alpha$-elements than the 
thin disk. This is also true for other elements as well, e.g. Eu and
Ba. From the, currently few, studies of the r- and the s-process elements
it is clear that the thick disk has experienced a contribution from AGB stars.

Some studies find that the thick disk has also experienced a contribution 
from SNe\,Ia. This is true in purely kinematically defined samples (e.g.
\cite{feltzing:bensby03}), but is not present, obviously, if the more metal-rich
stars, $>-0.5$ dex, are excluded as in \cite{feltzing:fuhrmann98}.

Based on currently available elemental abundance data and age 
determinations, the thick disk could have formed either through a 
violent, heating merger or through accretion of (substantial) 
satellites in a hierarchical galaxy formation scenario. The fast 
monolithic-like collapse is getting more and more problematic as data
are gathered. It would be especially crucial to establish if there
is an age-metallicity relation  in the thick disk or not as 
in that case the thick disk could not have formed in that way
(since the models indicate that the formation time-scale for the 
stars in  the thick disk would be very short, see \cite{feltzing:burkert}).

Thick disks are common in other galaxies, especially in merger environments, 
hence perhaps we should prefer the merger scenario for the Milky Way?

%

\end{document}